\documentclass[preprint,journal]{vgtc}       

\ifpdf
  \pdfoutput=1\relax                   
  \usepackage{graphicx}                
  \DeclareGraphicsExtensions{.pdf,.png,.jpg,.jpeg} 
\else
  \usepackage{graphicx}                
  \DeclareGraphicsExtensions{.eps}     
\fi%

\graphicspath{{figures/}{pictures/}{images/}{./}} 

\usepackage{microtype}                 
\PassOptionsToPackage{warn}{textcomp}  
\usepackage{textcomp}                  
\usepackage{mathptmx}                  
\usepackage{times}                     
\usepackage{cite}                      
\usepackage{tabu}                      
\usepackage{booktabs}                  

\usepackage{enumerate}

\usepackage[dvipsnames]{xcolor}

\usepackage{amsmath}
\usepackage{booktabs}
\usepackage{multirow}
\usepackage{graphicx}

\onlineid{1138}

\vgtccategory{Research}
\vgtcpapertype{Original Research}

\title{Action-Specific Perception \& Performance on a Fitts’s Law Task \\in Virtual Reality: The Role of Haptic Feedback}

\author{Panagiotis Kourtesis, Sebastian Vizcay, Maud Marchal, \textit{Member, IEEE}, \\ Claudio Pacchierotti, \textit{Senior Member,~IEEE}, Ferran Argelaguet}
\authorfooter{
\item
 P. Kourtesis, S. Vizcay, and F. Argelaguet are with Inria, Univ Rennes, IRISA, CNRS -- 35042 Rennes, France. E-mail: name.surname@inria.fr.
\item
 M. Marchal is with the Univ Rennes, INSA, IRISA, Inria, CNRS -- 35042 Rennes, France and Institut Universitaire de France.
\item
C. Pacchierotti is with the CNRS, Univ Rennes, IRISA, Inria -- Rennes 35042, France.
}

\shortauthortitle{Biv \MakeLowercase{\textit{et al.}}: Action-Specific P}
\shortauthortitle{Kourtesis \MakeLowercase{\textit{et al.}}: Action-Specific Perception on a Fitts's Law Task in VR}

\abstract{While user's perception and performance are predominantly examined independently in virtual reality, the Action-Specific Perception (ASP) theory postulates that the performance of an individual on a task modulates this individual's spatial and time perception pertinent to the task's components and procedures. This paper examines the association between performance and perception and the potential effects that tactile feedback modalities could generate. This paper reports a user study (N=24), in which participants performed a standardized Fitts's law target acquisition task by using three feedback modalities: visual, visuo-electrotactile, and visuo-vibrotactile. The users completed 3 Target Sizes $\times$ 2 Distances $\times$ 3 feedback modalities = 18 trials. The size perception, distance perception, and (movement) time perception were assessed at the end of each trial. Performance-wise, the results showed that electrotactile feedback facilitates a significantly better accuracy compared to vibrotactile and visual feedback, while vibrotactile provided the worst accuracy. Electrotactile and visual feedback enabled a comparable reaction time, while the vibrotactile offered a substantially slower reaction time than visual feedback. Although amongst feedback types the pattern of differences in perceptual aspects were comparable to performance differences, none of them was statistically significant. However, performance indeed modulated perception. Significant action-specific effects on spatial and time perception were detected. Changes in accuracy modulate both size perception and time perception, while changes in movement speed modulate distance perception. Also, the index of difficulty was found to modulate all three perceptual aspects. However, individual differences appear to affect the magnitude of action-specific effects. These outcomes highlighted the importance of haptic feedback on performance, and importantly the significance of action-specific effects on spatial and time perception in VR, which should be considered in future VR studies.

} 


\CCScatlist{ 
 \CCScat{K.6.1}{Management of Computing and Information Systems}%
{Project and People Management}{Life Cycle};
 \CCScat{K.7.m}{The Computing Profession}{Miscellaneous}{Ethics}
}

\teaser{
  \centering
  \includegraphics[width=\linewidth]{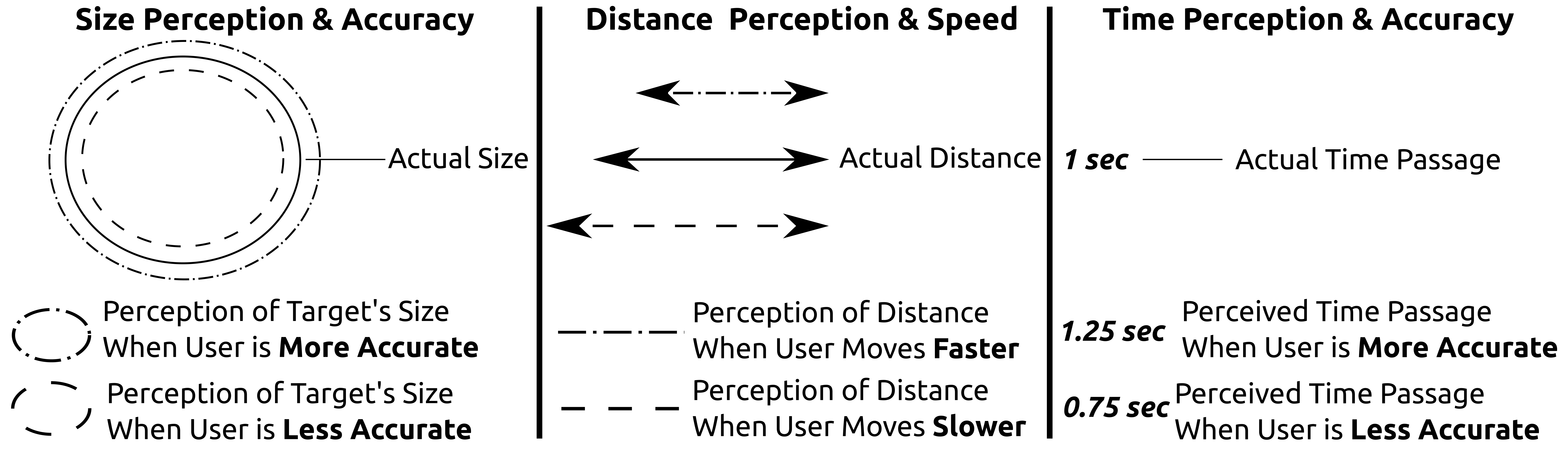}
  \caption{Illustration of the Action-Specific Effects on Size, Distance, and Time Perception.}
  \label{fig:teaser}
}

\vgtcinsertpkg

\begin{document}


\firstsection{Introduction}
\maketitle
Virtual reality is considered an effective tool to study perception \cite{Wilson2015} and performance \cite{Allcoat2018}. However, VR studies either examine perceptual processes (e.g., \cite{Jones2019,Suhaila2019,Zenner2017}) or performance-related aspects (e.g., \cite{Huawei2019,Khenak2020, Schwind2019}).  In contrast with this trend in VR studies, the Action-Specific Perception (ASP) theory, based on Gibson's ecological approach to perception \cite{Gibson2014}, postulates that the ability of an individual to perform a task modulates the individual's perception of this task's components and procedures  (see \autoref{fig:teaser}) \cite{Witt2011}. Furthermore, since the ability of an individual changes over time, due to various reasons, this individual's perception is altered correspondingly. However, the ability is not the only factor affecting perception. The difficulty of the task or required effort (e.g., climbing a hill with a heavy backpack and without it) also modulates perception \cite{Proffitt2003}. In VR, the only study that examined ASP is that of Laitin \emph{et al.,} \cite{Laitin2019}. However, this study examined only the effect of the observer's perceived effort/difficulty on distance perception, and it did not assess performance and/or interaction in VR (e.g., walking to the hill), nor other perceptual aspects (e.g., size or time perception). Thus, it has not been examined whether ASP is valid also in VR. If ASP's validity in VR would be confirmed (i.e., how well the user interacts with the virtual environment alters the user's perception of the virtual environment), it would affect how the components of and the interactions within the virtual environment should be designed, especially when an adaptive and/or personalized design is desired.%

In VR, haptic feedback devices have been used to improve user experience and performance \cite{Rakkolaine2021}. Passive haptic feedback has been found to increase performance on Fitts’s  law tasks \cite{Teather2010,Joyce2017}. Similarly, an electrovibration haptic interface/display was observed to improve performance on a Fitts’s  law task \cite{Zhao2020}. Though, passive haptic feedback and haptic displays are not preferred in VR, since VR applications require haptic devices with increased wearability and portability \cite{Maisto2017,Pacchierotti2017,Meli2014,Chinello2020}. In this respect, vibrotactile handheld devices are the most widely used haptic feedback devices in VR for improving user experience and performance \cite{Rakkolaine2021}.%

However, the vibrotactile has scarcely been examined against other haptic modalities in VR. Vibrotactile has mechanical elements which make it less wearable, more power-consuming, and slower \cite{Chourvardas,Kaczmarek,Pacchierotti2017}. An effective alternative to vibrotactile, could be the electrotactile feedback that is both portable and wearable, with lower power consumption and a high spatial and temporal resolution \cite{Chourvardas,Kaczmarek,KourtesisElec}. However, electrotactile and vibrotactile have not yet been compared in terms of user's performance in VR.%

Electrotactile feedback has been seen as a promising haptic medium for implementations in VR \cite{KourtesisElec}. However, its application in VR has not been extensively scrutinized. Pamungkas and Ward \cite{Pamungkas2016} used electrotactile feedback to render thermal (e.g., heat) and textural (e.g., roughness) tactile sensations. Hummel \emph{et al.,} \cite{Hummel} observed that electrotactile feedback substantially improved grasping accuracy. Finally, Vizcay \emph{et al.,} \cite{Vizcay2021} found that unimodal and bimodal (visuo-) electrotactile had a significant positive impact on contact accuracy. However, electrotactile has not yet been compared with vibrotactile, especially on a standardized target acquisition (Fitts’s law) task.%

While it is considered that vibrotactile improves performance, there are discrepant findings in the literature. Some studies reported that vibrotactile improves reaction time but not accuracy \cite{Burke2006}. Some other studies reported worse reaction time \cite{Bao,Kreimeier} or accuracy \cite{Cheng,Cholewiak} when using vibrotactile. Lastly, some studies indicated that there is no effect (positive or negative) at all \cite{vanBreda, Wang}. Also, vibrotactile feedback was found to facilitate worse performance than visual feedback \cite{Cheng,Cholewiak}. However, these discrepancies may have emerged due to the use of diverse non-standardized tasks with substantially different requirements.%

The implementation of a standardized task such as a Fitts’s law task would have enabled an examination of the performance benefits mediated from the use of each haptic feedback, as well as a scrutiny of any related action-specific effects on perception. Fitts’s law tasks are traditional target acquisition tasks in human-computer interaction (HCI) that can assess performance in terms of both speed and accuracy, and also consider the (objective) index of difficulty of the task \cite{MacKenzie1992}. While Fitts’s law was initially developed and examined in 2D spaces, its applicability in 3D spaces \cite{McGee1997,Murata2001} and VR HMDs \cite{Baekgaard2019,Clark2020} has been confirmed. Also, a Fitts’s law task is standardized by the ISO 9241-400:2007 \cite{ISO} for the ergonomics of human-system interaction.%

This paper offers a users' performance (accuracy and reaction time) comparison amongst visuo-electrotactile, visuo-vibrotactile, and visual-only feedback on a Fitts’s  law task in VR, and a scrutiny of the performance (accuracy and speed) and task's difficulty effects on size, distance, and (movement) time perception. The contributions of this paper and user study (N=24) can hence be summarized as follows:%

\begin{itemize}
    \item
    Conducted the first comparison of the effects of the electrotactile, vibrotactile, and visual-only feedback modalities on performance.%
    \item
    Conducted the first comprehensive examination of the ASP in VR, based on the performance on a Fitts’s law task.%
    \item
    Provided evidence that the ASP assessment is a more parsimonious method for evaluating the impact of haptic and input devices on performance.%
    \item
    Provided evidence that the user’s performance, concerning the interaction with the components of a virtual environment, affects how the user perceives the virtual environment.%
    \item 
    Provided a new utility of the index of difficulty of Fitts’s law for predicting and explaining action-specific effects on perception.%
\end{itemize}

\section{Relevant Work}
Fitts’s law is a predictive model of human movement that has been widely used in HCI and ergonomics \cite{MacKenzie1992}. Based on the Shannon–Hartley theorem for information transmission, Paul Morris Fitts \cite{Fitts1954}, and later, Scott Mackenzie \cite{MacKenzie1992} proposed an equation (see \autoref{FittsID}), where the width of targets ($W$) and the distance between targets ($D$) in a pointing selection task, define the required human movement in units of bits, which also resembles the task's index of difficulty ($ID$). The Fitts’s law task and metrics have been adopted by the ISO 9241-400:2007 of the International Organization for Standardization, regarding the ergonomics of human-system interaction \cite{ISO}. A Fitts’s law task hence is a standardized way to assess performance facilitated by various input devices and haptic devices. However, the electrotactile feedback has never been examined in a Fitts’s law task, and the vibrotactile has not been extensively examined in such tasks, albeit its extensive application in VR.%

\begin{equation}
ID = log_{2} \left( \frac{D}{W}+1 \right)
\label{FittsID}
\end{equation}%

\subsection{Performance \& Vibrotactile Feedback in VR}
In the literature, there are discrepant results regarding the performance benefits of vibrotactile feedback. In a meta-analysis of visuo-tactile feedback on human performance, the authors concluded that the addition of vibrotactile improves reaction time, but not accuracy \cite{Burke2006}. However, the meta-analysis also indicated the presence of several disagreements amongst the findings of diverse studies. In contrast with the meta-analysis conclusions, two studies found that the vibrotactile increases reaction time (i.e., slower) instead of improving it \cite{Bao,Kreimeier}. Also, the vibrotactile was found to enable a worse accuracy compared to the visual feedback \cite{Cheng,Cholewiak}. However, the discrepancies amongst the findings may be attributed to the use of diverse displays or tasks, with substantially different requirements.%

In contrast, Fitts’s law tasks are standardized tasks, which may provide more reliable inferences and conclusions. In a Fitts’s law task in non-immersive VR, the vibrotactile has been found to improve selection time compared to none feedback conditions (i.e., also an absence of visual feedback) \cite{Bora}. However, this study did not provide visual feedback to facilitate comparison against vibrotactile, and it was in normal screen display (non-VR). In contrast, using a Fitts’s law task in VR (stereoscopic display), the visual feedback facilitated a substantial better selection time than the visuo-vibrotactile condition \cite{Brickler}. The study that seems more proximal to ours, is the study of Ariza \emph{et al.,} \cite{Ariza}. In their study, no significant differences were detected among visual, vibrotactile, and visuo-vibrotactile conditions in performance on a Fitts’s law task in VR. However, significantly longer correction phases were in vibrotactile condition, which may indicate slower reaction times.%

\subsection{Performance \& Electrotactile Feedback in VR}
Electrotactile feedback involves the application of a mild electrical stimulation to the skin, that bypasses the skin's mechanoreceptors, and directly activates the nerve endings, which then elicit corresponding tactile sensations \cite{Chourvardas,Kaczmarek,KourtesisElec}. A local subdermal polarization by two or more electrodes produces an electrical current, which then activates the nerve endings. This electrotactile stimulation thus requires the placement of thin electrodes at the sensation point (e.g., the fingertip). Due to its reduced requirements (i.e., a single stimulator for multiple contact points, and the respective number of electrodes), compared to other haptic devices, the electrotactile devices have more potential to achieve enhanced wearability and portability.%

The literature pertinent to electrotactile feedback in VR is limited since the technology has recently re-emerged. Also, the electrotactile feedback has not been examined before in a Fitts’s law task. Previous implementations of electrotactile feedback in VR facilitated substantial improvements in users' performance \cite{KourtesisElec}. In a drilling teleoperation task in VR, the use of electrotactile feedback showed improved accuracy and speed \cite{Pamungkas2}. Similarly, in another robotic teleoperation in VR, the performance while using electrotactile feedback was comparable to the physical real-life control via exoskeleton \cite{Sagardia}. Also, in the follow-up study, where there was not any teleoperation involved, the electrotactile feedback assisted the users to have a more accurate grasping performance in VR \cite{Hummel}.%

Finally, the study of Vizcay \emph{et al.,} \cite{Vizcay2021} seems to be the closest to our study. In their study, the users performed a contact task in VR. The electrotactile feedback significantly improved the contact accuracy, while the visuo-electrotactile feedback facilitated the best accuracy. However, the performance benefits were detected only after the 2$^{nd}$ calibration of electrotactile feedback, suggesting that a suitable intensity of electrotactile feedback is required to observe performance benefits.%

\subsection{ASP, Performance, \& Task's Difficulty}
The ASP theory suggests that performance and the difficulty of a task modulate the perception of the task's components and procedures \cite{Gibson2014, Witt2011}. However, a crucial point is that the action-specific effects are significant only when there is an intention to act or the act has already been performed \cite{Witt2010}. In VR, we could say that the interactability of the components of a virtual environment is the prerequisite to observing action-specific effects on perception. Moreover, these action-specific effects have been repeatedly identified and confirmed in sports. More accurate soft-ball players perceived the ball as bigger \cite{Witt2005Ball}. Also, tennis players who were more efficient and accurate while returning the ball, perceived the ball's travelling time as slower \cite{Witt2010Tennis}. In hiking, hikers with a heavy backpack perceived the distance as longer \cite{Proffitt2003}. Note that the aforementioned action-specific effects are two-directional (e.g., less accurate soft-ball hitters perceived the ball as smaller).%

However, ASP is not valid only in sports. The action-specific effects on perception have also been detected in other populations and settings. Individuals who were weighted more \cite{SUGOVIC2016} and chronic pain patients \cite{witt2009long} perceived distances longer. Also, while playing a custom version of the videogame Pong, participants perceived the ball's movement faster when the paddle was smaller (i.e., increased difficulty) and they were less accurate, while when the paddle was larger and the participants were more accurate, then they perceived the ball's movement slower \cite{Witt2017}. However, this study used a non-standardized 2D task, where they manipulated only 2 levels of difficulty with large differences between them (i.e., the paddle's height was either 1.86 cm or 9.28 cm), and they investigated only size perception (of the paddle). Also, this study substantially differs from VR, where the interaction is facilitated directly by hand movements with 6 DoF. The only study of ASP study that was conducted in VR was that of Laitin \emph{et al.,} \cite{Laitin2019}. However, in this study, the VR users were just observers of hills. The walking distances over steep hills were perceived by the participants as longer. Nevertheless, this study examined only the effect of difficulty (i.e., steep hills) on distance perception of observers in VR, and it did not examine users' performance. Hence, this study did not investigate the user's performance or interaction with the virtual environment, and it did not provide evidence of whether the user's performance affects the user's perception of the virtual environment.%

Nevertheless, the over-confirmatory ASP findings and their replication has raised criticism that the observed action-specific effects on perception (or some of them) have been by-products of a response-bias \cite{FirestoneScholl2014,Laitin2019}. To address this criticism, the \emph{El Greco} evaluation method has been suggested for detecting findings mediated from response-biases \cite{FirestoneScholl2014,FirestoneScholl2016}. This method takes its name from the famous renaissance painter who used to draw thin and long figures in his paintings. The rationale is that if El Greco had a sight or brain-related issue, then he would also see the canvas elongated. So, if he had seen the real-world elongated, then he would have also seen his painting elongated, so the one effect would have cancelled the other effect, and the figures on his paintings would have been normally portrayed, and not elongated.%

A comparable logic has been implemented in ASP studies to evaluate action-specific effects. Hence, if an action-specific effect on a target object's perception is truly present, then it should also be present for the reference object. Consequently, after an initial evaluation of the action-specific effects on the target (e.g., asking participants to indicate the size of an object that they have previously interacted with), providing explicit feedback (i.e., a reference object) to the participants, by showing them the target object in its true form, would cancel the related action-specific effects, and then the participants would change their responses to match the reference object. If there is a significant difference between the initial evaluation and the El Greco evaluation, then this confirms the existence of action-specific effects on the perception of the target object. Inversely, a non-significant difference between the initial evaluation and the El Greco evaluation suggests the existence of response biases.%

\subsection{Summary}
Based on this literature review, the performance benefits of electrotactile and vibrotactile feedback in VR have not been previously compared, especially on a standardized task like a Fitts's Law target acquisition task. Also, the performance gains of using electrotactile feedback in VR have scarcely been investigated. Notably, electrotactile feedback effects on a Fitts's law task have not been previously evaluated. Similarly, the ASP has not been investigated comprehensively on one task (i.e., one performance affects the size, distance, and time perception). Notably, the ASP has not yet been examined whether it is valid in VR.  Moreover, the evidence supporting ASP is derived from either real-world cases that are considered extreme (e.g., professional sports players and disabled patients) or from extreme digital cases (e.g., tiny or large paddles, and steep hills). The VR users are not extreme cases (e.g., patients or professional players) and the interactable objects in the virtual environments do not have enormous size differences. Thus, the current study attempts to address all the above gaps by evaluating and comparing the performance facilitated by electrotactile and vibrotactile feedback on a Fitts's Law task; examining the ASP comprehensively (i.e., size, distance, and time) and whether the ASP is valid in VR; and investigating whether the ASP is valid in non-extreme populations and situations.%

\section{Experimental Methods}
This experiment will endeavour a comparison amongst visuo-electrotactile, visuo-vibrotactile, and visual feedback in terms of performance (accuracy, selection and reaction time) on a Fitts’s law task and action-specific effects on perception. Importantly, we aim to scrutinize the effects of performance and index of difficulty of Fitts’s law on size, distance, and time perception in VR.%

\subsection{Fitts’s Law Task}
\subsubsection{User-Centred Design}
\begin{figure}[!h]
 \centering 
 \includegraphics[height = 280pt]{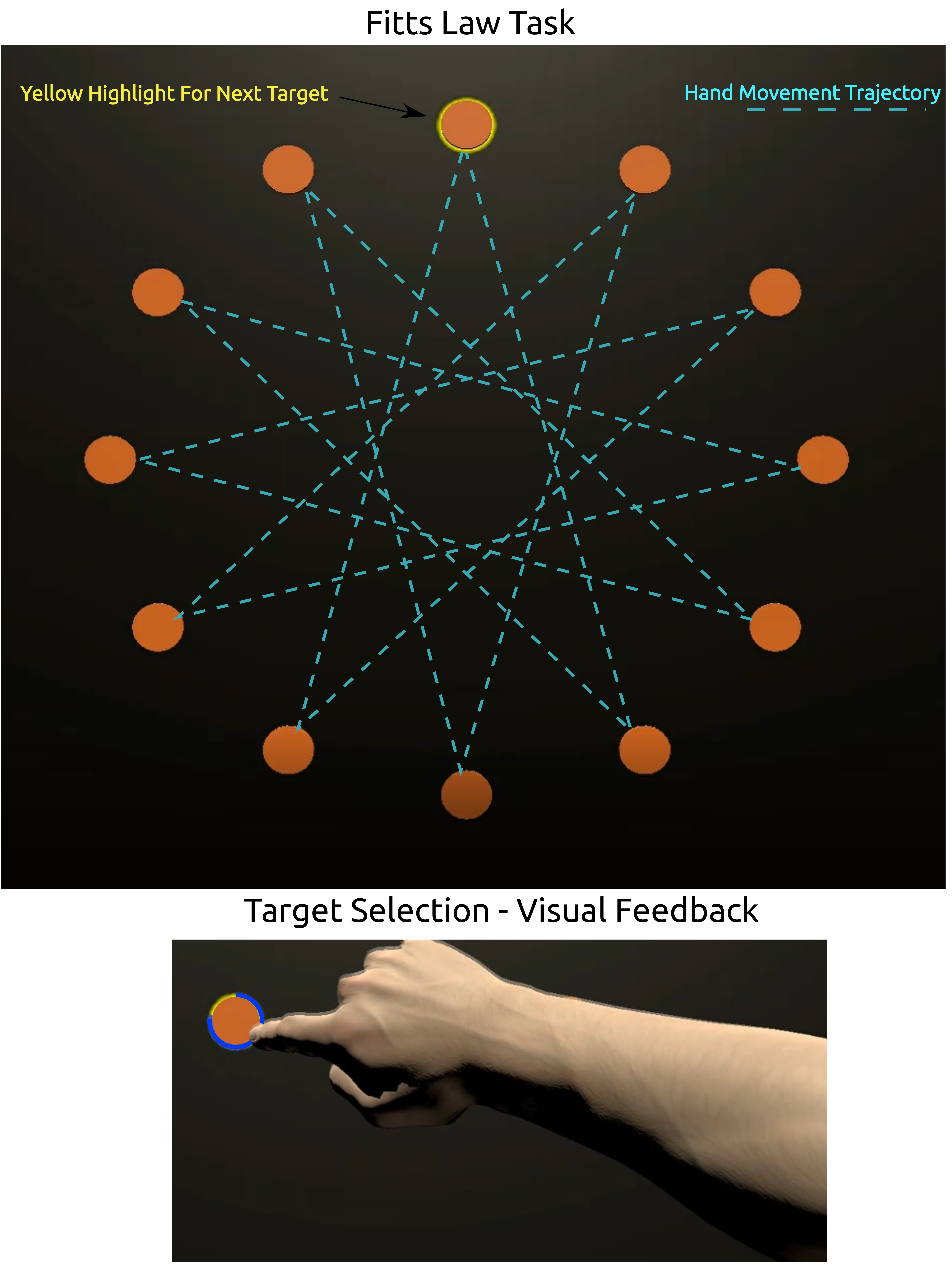}
 \caption{Fitts’s Law Task (Top) and Visual Feedback (Bottom).}%
 \label{fig:VisTask}
\end{figure}

Gaming and/or computing ability may affect the performance of users. However, this has predominantly been seen in controller-based interaction. In contrast, ergonomic and naturalistic interactions have been found to significantly mitigate the effects of gaming ability on performance, facilitating a comparable performance between gamers and non-gamers \cite{Kourtesis2021}. Consequently, we opted for an ergonomic and naturalistic interaction for performing the selection task, where the user has a virtual hand with a predefined pose (i.e., a slightly firm fist with the index finger pointing; see \autoref{fig:VisTask}) that is being collocated with user's physical hand by using a Vive tracker v2.0 placed on their wrist.%

In selection tasks, an important point of consideration is the ``Heisenberg effect'', which suggests that a confirmation action (e.g., pressing a button on the controller) simultaneously with the pointing action (i.e., maintaining the pointer on the target) may affect the accuracy of the selection \cite{Argelaguet2013}. To avoid a ``Heisenberg effect'', the confirmation in the selection task is automatic. The user is just required to point at the target for a predefined time (i.e., a second) to select the target.%

Furthermore, another factor that may affect performance is the occlusion of the target, which especially occurs in cluttered environments \cite{Argelaguet2013}. To address this issue, we ensured that the field of view from the seated position of the users is well beyond the task area. Hence, the user always sees all the 12 targets. Also, the smallest target's width was defined to be 2 cm to avoid being occluded by the virtual index finger. Finally, to avoid confusion about the next target, we used a yellow highlight (see \autoref{fig:VisTask}) that informs the user which the next target is.%

Additionally, the virtual hand representation may affect both the sense of presence and performance \cite{Kilteni2013}. A realistic virtual hand that is also well collocated with the physical hand can elicit a strong sense of hand agency and ownership \cite{Argelaguet2016}. However, the hand's gender should match the user's gender \cite{Schwind2017}. Correspondingly, we used two realistic virtual hands, one for each gender, which were collocated (see mention to tracker above) with the user's hand. Also, there was a left or right virtual hand for each gender to match their dominant hand. Finally, to consolidate realism, the virtual hands could not interpenetrate the virtual objects (i.e., they were colliding, with respective hand-bending animations).%

Finally, the use of color-blind friendly colors addresses accessibility issues pertinent to color-blind individuals, as well as facilitates a further improvement of the ease of access to non- color-blind individuals that may ameliorate their performance \cite{Ichihara2008}. For this target selection task, we used 4 colors from the Okabe-Ito palette, which have previously been found effective for providing a color-blind friendly discrimination amongst them \cite{Ichihara2008}. Thus, black was used for the background of the task, orange for the targets, blue for the visual feedback (i.e., a loading circle around the target, indicating that it is being selected), and yellow for the highlight circle indicating the next target (see \autoref{fig:VisTask}).%

\subsubsection{Task Design \& Index of Difficulty}
The task was designed by following the guidelines of ISO 9241-400:2007 \cite{ISO}. Correspondingly, the targets were placed in a vertical plane to offer the optimal ergonomics for seated participants. The Fitts’s  law task included 12 targets placed in a circular form with equal distances, from one to the next target. The hand movement trajectory for selecting all the targets was forming a star-like shape (see \autoref{fig:VisTask}). The user had to select each target twice, thus, the user had to complete two rounds in each trial. The targets in each trial had a different width. There were 3 target sizes: 1) small (diameter = 2 cm); 2) medium (diameter = 3 cm); 3) large (diameter = 4 cm). Also, the distance between the targets had two different lengths: 1) Short = 40 cm; 2) Long = 80 cm. Using the Fitts’s  Law equation (see \autoref{FittsID}) \cite{Fitts1954}, we calculated six (3 target sizes $\times$ 2 distances) indexes of difficulty, one for each combination of target size with distance length. The calculated IDs are: \emph{Long-Small = 4.4}; \emph{Long-Medium = 4.1}; \emph{Long-Large = 3.8}; \emph{Short-Small = 3.5}; \emph{Short-Medium = 3.2}; \emph{Short-Small = 2.9}.%

\subsection{Feedback Design}
\subsubsection{Tactile Feedback Devices}
For the electrotactile feedback, we used a multichannel electrical stimulator with an electrode placed at the finger pad of the index finger of the dominant hand. The communication with the PC was made through an Bluetooth 3.0 at 115.2 bps. The electrical stimulation consists of square pulses with a pulse width of 500 $\mu$s and being delivered at a frequency of 200 Hz. This electrotactile device has effectively been implemented in other studies (e.g., \cite{Isakovic2022,Vizcay2021}).%

For the vibrotactile feedback, we used a standard commercial 5mm vibration motor \cite{vibrator304-116} capable of vibrating at a frequency between 120 and 280 Hz, which was also placed at the finger pad of the dominant hand and controlled (vibration's amplitude) by an Atmega 328 microcontroller attached on the Arduino board. The communication with the PC was through an RN42 Bluetooth module (v 3.0) connected to an Arduino mini pro 3.3 V. This vibrotactile device has efficiently been implemented in other studies (e.g., \cite{Designe2020,Cehajic2017}).%

\subsubsection{Dynamic vs Binary Feedback}
The electrotactile device has been previously used to offer dynamic feedback proportionally to the contact's accuracy \cite{Vizcay2021}. So, we intended to implement a similar mechanism, where the feedback's intensity would be stronger when the contact point (with the target) is closer to the target's centre. Also, the visual-feedback (i.e., a blue loading circle around the target; see \autoref{fig:VisTask}) could offer dynamic feedback, where the speed (0-1) of loading was analogous to the distance from the centre. We thus conducted a pilot study (N=5) to explore whether the vibrotactile can offer dynamic feedback comparable to the electrotactile one. Unfortunately, all participants reported an effortless identification of latency among the various amplitude changes. This noticeable delay was attributed to the mechanical parts of the vibrotactile device that require more time to change the vibration's frequency (i.e., strength).%

However, there was no perceived latency regarding a binary mode (i.e., activation/deactivation) of feedback. For this reason, we opted to offer a binary mode of feedback for every type (visual, electrotactile, and vibrotactile), where the feedback is either activated or deactivated, to provide a comparable tactile sensation, which will allow a fairer comparison between them. The results of our pilot and the preference for a binary type of feedback are in line with the findings of Ariza \emph{et al.,} \cite{Ariza}, where the binary vibrotactile feedback enabled a substantially better performance on a Fitts’s law task, compared to the dynamic vibrotactile feedback. Consequently, in our study, the feedback is provided (activated) when the user's index finger is colliding with the target until the selection of the target. At the selection of the target, the target disappears, which automatically deactivates all types of feedback.%

\subsubsection{Calibration}
\begin{figure}[!h]
 \centering 
 \includegraphics[width=\columnwidth, height = 120pt]{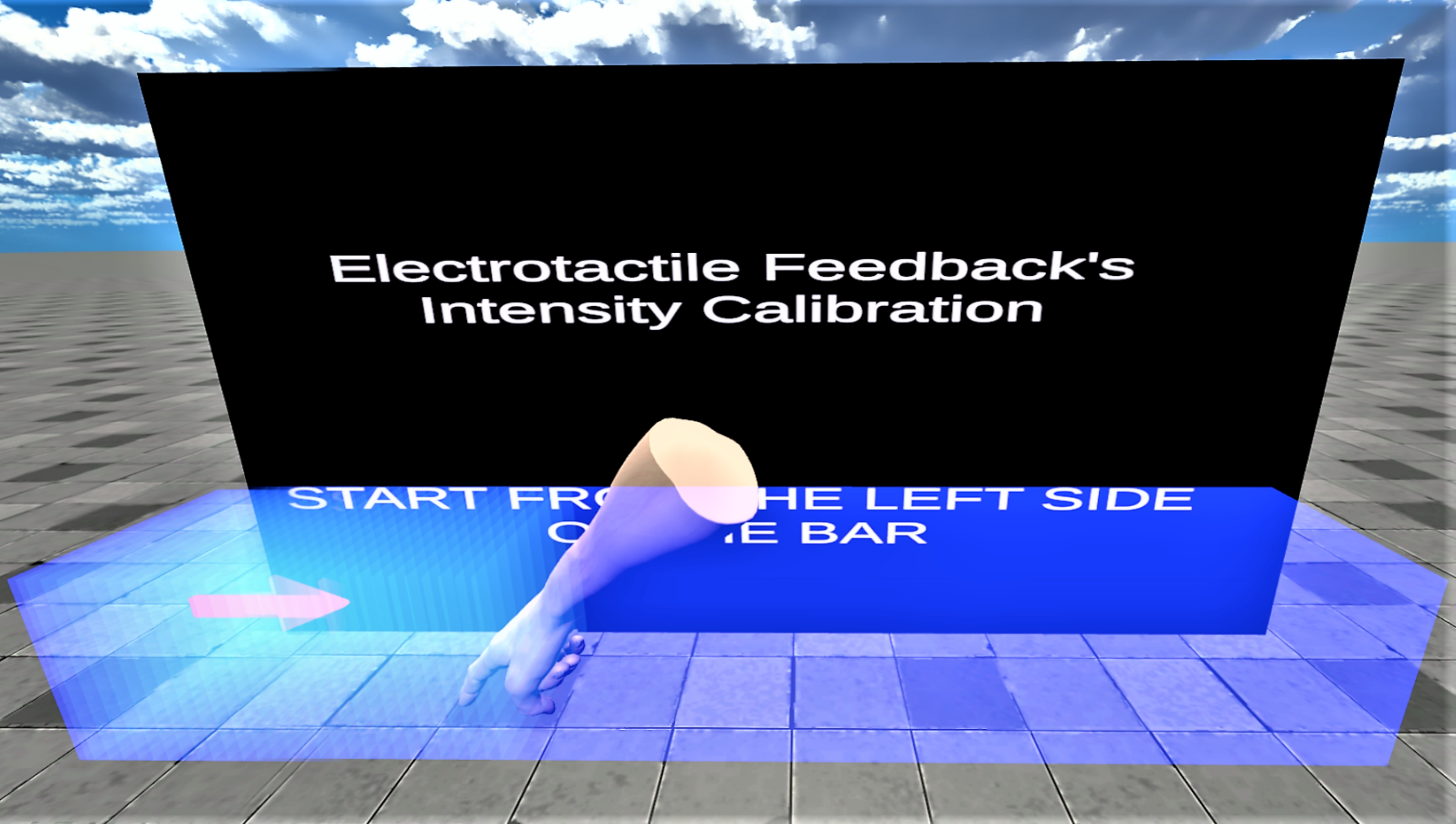}
 \caption{VR Calibration System for Electrotactile and Vibrotactile.}%
 \label{fig:Calib}
\end{figure}

The amplitude of the electrical stimulation was determined by the participant via a calibration procedure to ensure a perceivable and not unpleasant sensation. 
To guarantee an optimal intensity of the electrotactile feedback during the task's performance, the calibration was performed twice, once at the very beginning of the VR exposure (initial calibration), and then just before the electrotactile condition (actual calibration). 
During calibration, participants could set the intensity to any value between 0.1 and 9.0 mA.
Since the vibrotactile is a very familiar sensation (e.g., extensively used on smartphones and tablets), it was calibrated just once. Comparably to the electrotactile feedback, just before the vibrotactile condition, participants performed a calibration to determine the strength of the vibrotactile feedback that they prefer. The calibration of electrotactile and vibrotactile was performed in VR by interacting with a VR slider bar (see \autoref{fig:Calib}), where movement towards the right side was increasing the intensity, and towards the left side was decreasing it.%

\subsection{ASP Evaluation}
\begin{figure}[!b]
 \centering 
 \includegraphics[width=\columnwidth]{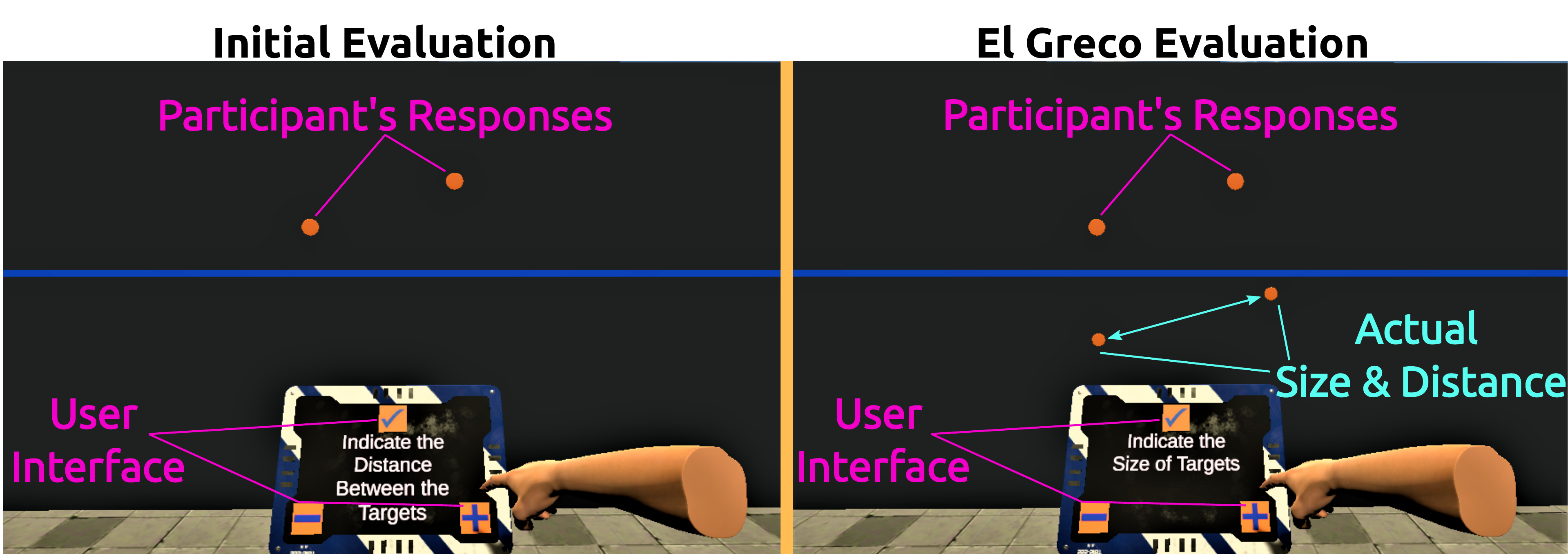}
 \caption{ASP: Initial Evaluation (Left) \& El Greco Evaluation (Right)}%
 \label{fig:ASP_Task}
\end{figure}

The action-specific effects were assessed directly after the completion of the Fitts’s law task in each trial. The participants by using a plus, a minus, and a confirmation button, could correspondingly increase, decrease, and confirm their responses (see \autoref{fig:ASP_Task}). The participants had to indicate the target size, the distance between the targets, and the time required to go from one target to the other.%

First, the participants had to change and indicate the size of two targets with a starting diameter = 1cm. In each press, the plus and minus buttons changed the size (diameter) of both targets by 0.1cm. Secondly, the participants had to change and indicate the distance between the targets, with a starting distance = 10cm. In each press, the plus and minus buttons increased the distance by 1cm. Lastly, the participants had to change and indicate the required movement time, with a starting time at 0s. With each press, the plus and minus buttons changed the time by 0.1s. Note that the participants could keep the plus or minus button pressed to increase or decrease the values faster. After releasing, by single presses, they could specify their responses, and finally, they could press the confirm button to validate them.%

At the end of each trial, this procedure was completed twice, once for the initial evaluation, and once for the El Greco evaluation. The initial evaluation was performed immediately after the completion and disappearance of the Fitts’s law task. In the initial evaluation, the participants did not have any visual cues or feedback about the size of the targets and the distance between them.
After the initial evaluation, the participants performed the El Greco evaluation, which was destined to cancel any action-specific effects by providing explicit feedback on the target size and distance between the targets. During the El Greco evaluation, the participants could see the actual target size and distance between them, under the blue line (see \autoref{fig:ASP_Task}). Besides the display of the reference targets for informing the participants about the actual target's size and distance between them, the procedure of the El Greco evaluation was identical to the initial one.%

\subsection{Participants \& Procedures}
\paragraph{Participants:}All participants signed an informed consent form to participate in this study. A meticulous hygiene protocol (use of anti-viral gels and sprays, cleaning of apparatuses and HMDs, use of protective covers, and wearing masks) was followed to ensure augmented hygiene standards due to the COVID-19 pandemic. The study followed Helsinki guidelines, and it was approved by the institution's ethical committee. We recruited a total of 24 participants (10 Females, 14 Males) with an age (years) $mean(SD) = 27.12 (4.41), range = 21-39$, an education (years) $mean(SD) = 17 (1.66), range = 14-21$. The participants had a frequency of using VR ($range = 1-6$) $mean(SD) = 3.12(1.88)$ (3 = A few times a month), and a self-reported ability in using VR  ($range = 1-6$) $mean(SD) = 3.63(1.54)$ (3 = Average, 4 = Skilled). Their experience on VR (i.e., frequency + ability) was ($range = 2-12$) $mean(SD) = 7(3.34)$. Based on the VR experience score, 13 participants were above the median of 6.5 (i.e., skilled), and 11 participants were below it (i.e., low skills or no experience). Likewise, the participants had a gaming frequency ($range = 1-6$) $mean(SD) = 3.46(2.11)$, ability ($range = 1-6$) $mean(SD) = 3.50(1.51)$, and an overall gaming experience ($range = 2-12$) $mean(SD) = 7.18(3.42)$. Based on the gaming experience score, 14 participants were skilled (i.e., gamers above the median score of 6.5) and 10 were poorly skilled (i.e., non-gamers) below the median.%

\begin{figure}[!h]
 \centering 
 \includegraphics[width =\columnwidth]{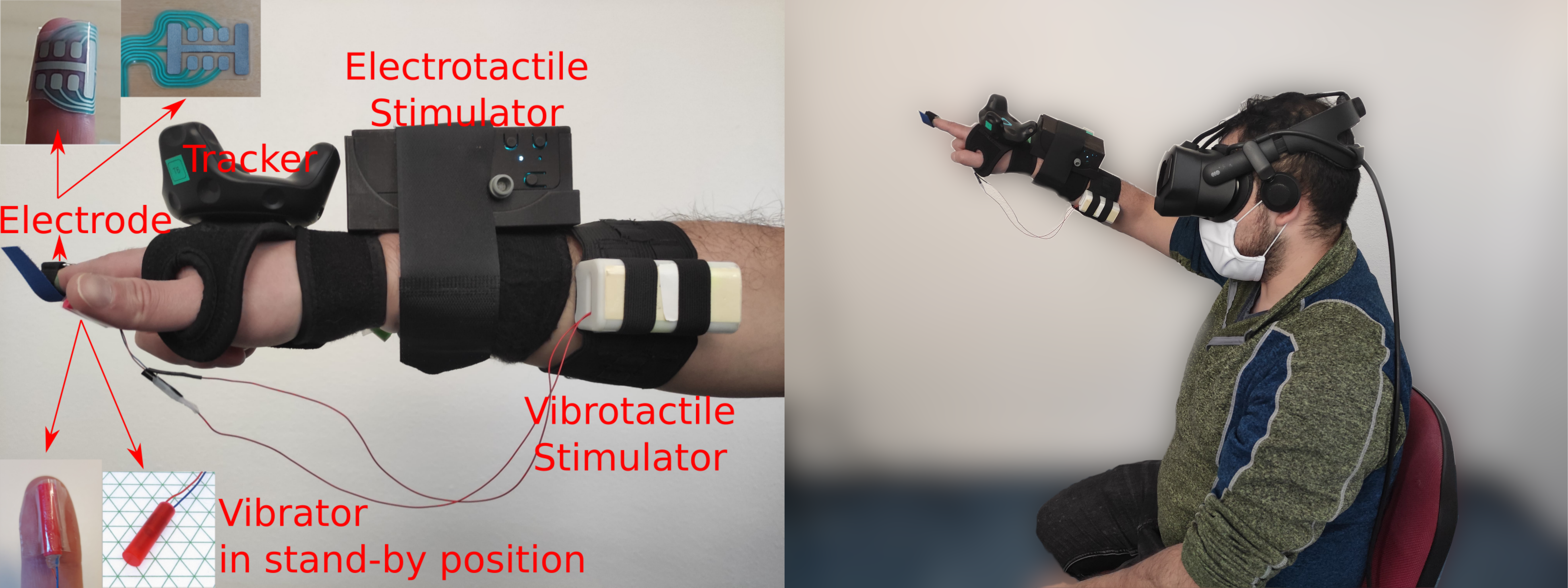}
 \caption{Experimental Set Up: Haptic Devices \& Tracking (Left) and Participant Performing the Fitts’s Law Task (Right).}%
 \label{fig:SetUP}
\end{figure}

\paragraph{Set Up:} We followed the ISO 9241-400:2007 \cite{ISO} for offering an ergonomic interaction for seated participants. The distance between the seated position of the participant to the Fitts’s Law or ASP task was 50 cm. The height of the centre of the tasks was 110 cm to approximate the head's height in a seated position (see \autoref{fig:SetUP}). We used a valve index HMD with v2 lighthouse stations to immerse and track the participants' dominant hands and heads. A v2 Vive tracker was placed on the wrist of participants to track their hand and collocate it with the virtual hand. The electrotactile stimulator was attached to the participants' forearms by using an armband. Similarly, the vibrotactile actuator was attached to the forearm (behind the electrotactile stimulator) by using another armband (see \autoref{fig:SetUP}). The participants had both the electrotactile stimulator and the vibrotactile actuator attached to their forearm in all conditions to mitigate any confounding factors pertinent to ergonomics. By using tape, the electrode and the vibro-motor were attached either to the index finger pad (i.e., distal phalanx) or the thumb's finger pad, based on the experimental block. For example, in the vibrotactile condition, the vibro-motor was attached to the index and the electrode to the thumb. Also, an adjustable Velcro was used during the electrotactile condition, to maintain a mild yet constant pressure, and ensure good contact between the skin and the electrode.%

\subsubsection{Procedure}
The experiment was developed in Unity 3D. Amazon Polly’s neural text-to-speech was used for creating audio files for the verbal instructions in VR. Before getting immersed, the participants filled out a questionnaire with their demographic data, such as age, sex, educational level, frequency of and ability in playing videogames, and frequency of and ability in using VR. Also, the experimenter described the experimental procedures to the participants and emphasized the importance of performance in terms of both speed (moving fast from one target to the other) and accuracy (aiming at the centre of the target). In VR, the participants firstly calibrated the electrotactile feedback (initial calibration). Then, the participants went through an audio-visual (i.e., video and verbal instructions) tutorial about the Fitts’s  Law task and ASP task. The participants also completed a practice trial for each task.%

There were 3 blocks, one for each feedback type: visual-only, visuo-electrotactile, visuo-vibrotactile. At the beginning of each block, the experimenter was placing the respective tactile compartment (i.e., electrode or vibro-motor) at the participant's index pad, before allowing (pressing the space bar) the participant to continue. During the placement of the electrode, the experimenter was checking whether there is moisture (i.e., sweat), to wipe it accordingly. Then, the participant could proceed with the calibration (actual) of the tactile feedback (for electro- and vibrotactile), followed by a practice trial (only for the Fitts’s  Law task) specific to the block's type of feedback.%

Each block was comprised of 3 target sizes $\times$ 2 distances = 6 trials, where in each trial the participant performed the Fitts’s  Law task, and then the ASP evaluation. Before each trial, a written (label) and a verbal (audio) reminder was given to the participants to be both fast and accurate. Also, the participants were reminded that could rest as much as they want since the trial was commencing only when they touch the 1$^{st}$ target. In total, the participants completed \emph{3 target sizes $\times$ 2 distances $\times$ 3 feedback conditions = 18 trials}. In each trial, the participants selected 12 targets $\times$ 2 rounds = 24. Hence, each participant made 24 $\times$ 18 = 432 selections in the whole experiment.%

After the VR exposure, the participants filled out an evaluation questionnaire (7-point Likert scale) regarding the types of feedback (visual, electrotactile, and vibrotactile) in terms of pleasantness, helpfulness (overall, speed, accuracy), alertness, salience, and coherence. Specifically, the questions were \emph{"Please rate how much  X was each type of feedback:"} (X = \emph{pleasant}, \emph{alarming}, \emph{salient}, or \emph{coherent with your actions}); \emph{"Please rate how much helpful was each type of feedback for X :"} (X = \emph{performing EFFECTIVELY the task}, \emph{being ACCURATE in the task}, or \emph{being FAST in the task}). The available responses were respectively \emph{Not at All X}, \emph{A little bit X}, \emph{Somewhat X}, \emph{X}, \emph{Very X}, \emph{Very much X}, and \emph{Extremely X}. Also, they evaluated the dominance of vibrotactile and electrotactile against visual feedback. The questions were \emph{Please rate how much dominant was X against Visual Feedback:} (X = \emph{Vibrotactile} or \emph{Electrotactile}). The available responses ranged from \emph{Visual Feedback was Totally Dominant} to \emph{X was Totally Dominant}. Then, the participants responded to the NASA Task Load Index questionnaire, which indicates perceived workload in terms of mental, physical, temporal, performance, and effort demands \cite{Hart2006}. Finally, the cybersickness symptoms were assessed pre- and post- VR exposure, by using the CSQ-VR questionnaire, which has shown a very good structural validity \cite{Kourtesis2019}, and a robust convergent validity against SSQ, while it is faster to administer with more interpretable scores \cite{Somrak2021}. None of the participants experienced cybersickness symptoms.%

\subsection{Variables \& Hypotheses}
The variables were selected based on both their generalizability and their relationship with ASP theory.%

\paragraph{Performance:}For performance, we considered the accuracy index, the selection time, and the reaction time. The selection time is the time required from the moment that the target is selected until the selection of the next target. The reaction time is the time required to accelerate and leave the area of the target that was selected. The accuracy index was calculated based on the effective width of Fitts’s  law equation \cite{Fitts1954}. The average distance from the target centre ($\overline{DC}$), while interacting with it, was subtracted from the effective width ($W_e$) to calculate the accuracy index ($AI$; see \autoref{AccIndex}). A higher accuracy index hence indicates better accuracy, and vice versa, while it can also have negative values (e.g., when the average distance is greater than the effective width). Finally, the independent variable for performance was the feedback type.%

\begin{equation}
\begin{aligned}
W_e = 4.133 \; \times \; SD_{DC} \\
AI = W_e \; - \; \overline{DC}
\end{aligned}
\label{AccIndex}
\end{equation}%

\paragraph{ASP:}For ASP, we considered the difference between perceived (initial evaluation) and actual size, distance, and (average) movement time in each trial. Hence, we ended up with 3 dependent variables pertinent to ASP, respectively. Regarding the independent variables, they were the feedback type, the index of difficulty from Fitts’s law, the movement time (average time to just go from one target to the next one), and the accuracy index.%

\paragraph{Hypotheses:} Based on the relevant studies, we formulated the following hypotheses:
\begin{itemize}
    \item \textbf{[H1]} Visuo-electrotactile feedback will facilitate a significantly better accuracy than the visual and the vibrotactile, while the speed will be comparable to the visual feedback condition. 
    \item \textbf{[H2]} Visuo-vibrotactile feedback will offer a comparable accuracy to the visual feedback, while the reaction time will be slower than the visual condition.
    \item \textbf{[H3]} The performance advantages attributed to each haptic feedback will be reflected in the action-specific effects on size, distance, and time perception.  
    \item \textbf{[H4]} Accuracy will affect both size and time perception, while speed (movement time) will affect distance perception. 
    \item \textbf{[H5]} Index of difficulty will affect size, distance and time perception. 
\end{itemize}

\section{Results}
The normality assumption of the distribution of DVs was examined. Both performance and ASP variables violated the assumption check. The data were thus transformed and centralized by using the $bestNormalize$ R package \cite{Peterson2020}. The DVs were then normally distributed. One-way Repeated-Measures ANOVA analyses were performed to explore the effect of the type of haptic feedback on the DVs. The  Greenhouse  Geisser’s correction was applied in the cases where the sphericity assumption was violated. Post-hoc comparisons were then executed for exploring the potential differences among them. The Bonferroni correction was used to amend p-values’ inflation due to multiple comparisons. The Hedge's $g$ and the $\mathcal{\omega}^2$ are reported since they are more parsimonious and reliable metrics of effect sizes for the ANOVA analyses and posthoc comparisons, respectively \cite{Lakens2013}.%

The self-reports on each feedback's utility were compared by using the Wilcoxon Rank Sum Test. The relationships between DVs, self-reports, and demographics were assessed by conducting Pearson's R and Spearman's Rho correlational analyses for continuous and ordinal variables correspondingly.  Also, the ASP was checked by comparing the initial and the El Greco estimation. The ASP aspects were further scrutinized by Pearson's R and multivariate regression analyses. The multivariate regression analyses were evaluated in terms of $\mathcal{\chi}^2$ probability for significance, and the $R^2$ for effect size (i.e., percentage of the variance explained by the model). Since the regression models are pertaining to perceptual processes and effects, the interpretation of $R^2$ was based on the recommendations of Cohen \cite{Cohen1992} for psychological sciences. 
The  $afex$ (ANOVA analyses;\cite{afex}), $ggstatsplot$ (plots, t-tests, and post hoc comparisons;\cite{ggplot}), and $lme4$ (regression analyses;\cite{lme4}) R packages were used.%

\subsection{Performance}
\begin{figure}[!h]
 \centering 
 \includegraphics[width=\columnwidth]{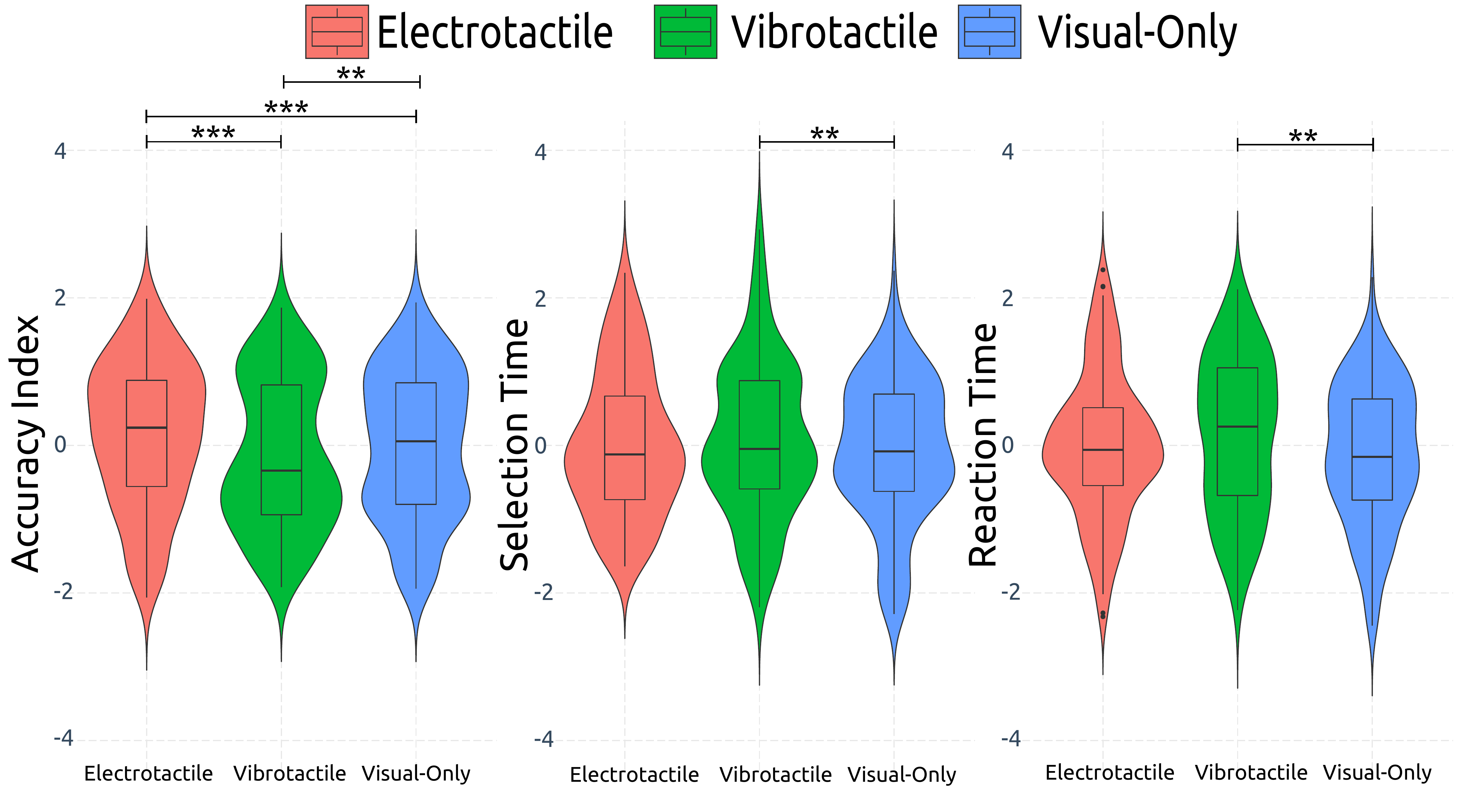}
 \caption{Comparisons between the Types of Feedback in terms of \\Accuracy (Left), Selection Time (Centre), and Reaction Time (Right). \\*** $<.001$, **$<.01$, and *$<.05$}%
 \label{fig:perf}
\end{figure}

\paragraph{Accuracy Index:} 
In support of \textbf{H1} \& \textbf{H2}, the main effect of the type of feedback on accuracy was statistically significant $F(1.99,284) = 25.16$, $p <.001$ indicating a substantial difference amongst the feedback types. However, the effect size was small $\mathcal{\omega}^2 = .010$, $95\%$ $CI[0,1]$. In support of \textbf{H1}, the post-hoc comparisons revealed significant differences between the feedback types, where electrotactile feedback appeared to facilitate substantially better accuracy (see \autoref{fig:perf}). Compared to Visual-only, the electrotactile feedback had a substantially improved accuracy, though, with a small effect size $g = .16$, $p <.001$, Comparably, against the vibrotactile, the electrotactile feedback facilitated a significantly better accuracy, with a small effect size $g = .27$, $p <.001$. Interestingly, yet in disagreement with \textbf{H2}, the visual-only showed significantly better accuracy than the vibrotactile, while the effect size was small $g = .12$, $p =.01$.%

\paragraph{Selection \& Reaction Time:} 
The ANOVA analyses surfaced significant main effects for both selection time  $F(1.90,272) = 3.77$, $p =.03$ and reaction time $F(1.76,251) = 4.30$, $p =.02$. However, the effect sizes were very small for both selection  $\mathcal{\omega}^2 = .003$, $95\%$ $CI[0,1]$ and reaction time  $\mathcal{\omega}^2 = .008$, $95\%$ $CI[0,1]$. In line with \textbf{H1}, the post-hoc comparisons (see \autoref{fig:perf}) showed that the electrotactile feedback did not significantly differ from visual-only and vibrotactile feedback conditions. However, the reaction and selection time in the electrotactile feedback condition were closer to the visual-only rather than the vibrotactile condition. In agreement with \textbf{H2}, compared to the visual-only condition, the selection time $g = .14$, $p =.01$. and the reaction time $g = .25$, $p =.003$. were both substantially slower in the vibrotactile condition (see \autoref{fig:perf}). However, the effect sizes of these differences were small.%

\subsection{Demographics \& Self-Reports}
The demographics (age, sex, education) did not show any correlation with the performance or the ASP variables. Similarly, the ability, frequency, and experience in playing video games or using VR did not correlate with any of the performance or the ASP variables. Also, none of the weighted scores of NASA-TLX (i.e., perceived workload) was correlated with the performance or the ASP variables. Equally, the intensities of electrotactile feedback and vibrotactile feedback did not show any correlation with the performance. 

\paragraph{Self-Reports on Feedback's Utility:}
The self-reports on electrotactile and vibrotactile indicated that they helped the participants to perform better. However, only the vibrotactile sensation was found to be pleasant, while the electrotactile and visual were found to be "somewhat pleasant". Nonetheless, no significant differences were detected between vibrotactile and electrotactile. Significant differences were identified only against visual feedback. In \autoref{tab:SelfReports}, there are displayed the descriptive data and the significant differences against visual feedback. Aligned with \textbf{H1}, compared to visual only, the electrotactile feedback was reported that significantly assisted the participants to be accurate. However, in disagreement with \textbf{H2}, compared to visual, the self-reports indicated that the vibrotactile assisted significantly more than participants with being accurate. Discrepantly to our \textbf{H2}, the self-reports did not reveal a significant difference in terms of helpfulness for performing fast the task.%

Furthermore, both the electrotactile and vibrotactile were evaluated as substantially more alarming than visual feedback, postulating that the two haptic feedback systems assist with increasing the vigilance of the user. Comparably, the two haptic feedback systems were reported to be significantly more salient than the visual feedback. However, all types of feedback were equally coherent (to very coherent). Lastly, there was no significant difference between electrotactile and vibrotactile in terms of dominance against the visual feedback. However, the vibrotactile was reported as more dominant than visual feedback, while the electrotactile feedback was reported as marginally more dominant (i.e., the median between \textit{neither}/\textit{both} and \textit{more dominant}).%

\begin{table}[!h]
\resizebox*{\columnwidth}{1.5in}{%
\fontsize{14pt}{22pt}
\selectfont
\begin{tabular}{@{}ccccccccc@{}}
\toprule
\multicolumn{1}{c|}{} & \multicolumn{2}{c|}{Visual} & \multicolumn{3}{c|}{Vibrotactile}       & \multicolumn{3}{c|}{Electrotactile}     \\ 
\multicolumn{1}{c|}{} &
  \multicolumn{1}{c|}{Median} &
  \multicolumn{1}{c|}{Range} &
  \multicolumn{1}{c|}{Median} &
  \multicolumn{1}{c|}{Range} &
  \multicolumn{1}{c|}{p-value} &
  \multicolumn{1}{c|}{Median} &
  \multicolumn{1}{c|}{Range} &
  \multicolumn{1}{c|}{p-value} \\ \midrule
Pleasantness          & 3               & 1-5            & 4   & 1-6 & .09                         & 3   & 0-5 & .43                         \\
Helpfulness-Overall   & 4               & 1-6            & 5   & 2-6 & .24                         & 5   & 2-6 & .33                         \\
Helpfulness-Accuracy  & 3               & 0-6            & 4   & 0-6 & \textbf{.03*}               & 4   & 0-6 & \textbf{.04*}               \\
Helpfulness-Speed     & 3               & 1-6            & 4   & 1-6 & 1                           & 3.5 & 1-6 & .59                         \\
Alarming              & 2               & 0-6            & 4.5 & 2-6 & \textbf{\textless{}.001***} & 4.5 & 3-6 & \textbf{\textless{}.001***} \\
Salience              & 3               & 1-6            & 5   & 4-6 & \textbf{.002**}             & 5.5 & 2-6 & \textbf{.005**}             \\
Coherence             & 4               & 2-6            & 4.5 & 1-6 & 1.00                        & 5   & 2-6 & 1.00                        \\
Dominance             & \textit{NA}     & \textit{NA}    & 4   & 0-6 & \textit{NA}                 & 3.5 & 0-6 & \textit{NA}     \\ \cmidrule(l){0-8}         
\end{tabular}%
}
\caption{Self-Reports on each Feedback's Utility. Maximum Score = 6; p-values are related to the comparison against Visual feedback.}
\label{tab:SelfReports}
\end{table}%

\subsection{Action-Specific Perception}
For the target size perception, compared to the initial estimation $mean(SD)=-0.06(0.56)$,  the El Greco estimation $mean(SD)=0.03(0.14)$ was significantly closer $g = .20$, $p <.001$ to the actual target size. For the targets' distance perception, compared to the initial estimation $mean(SD)=3.37(10.86)$,  the El Greco estimation $mean(SD)=0.09(3.36)$ was substantially closer $g = -.30$, $p <.001$ to the actual distance between the targets. Lastly, regarding the movement time perception, the comparison between El Greco $mean(SD)=0.01(0.37)$ and initial $mean(SD)=0.04(0.48)$ estimation showed that the El Greco was substantially closer $g = .18$, $p =.008$ to the actual movement time required to go from one target to the other. These results confirmed that there were indeed action-specifics effects on perception during the initial estimation, which were then efficiently cancelled by the El Greco estimation.%

\begin{figure}[!h]
 \centering 
 \includegraphics[width=\columnwidth]{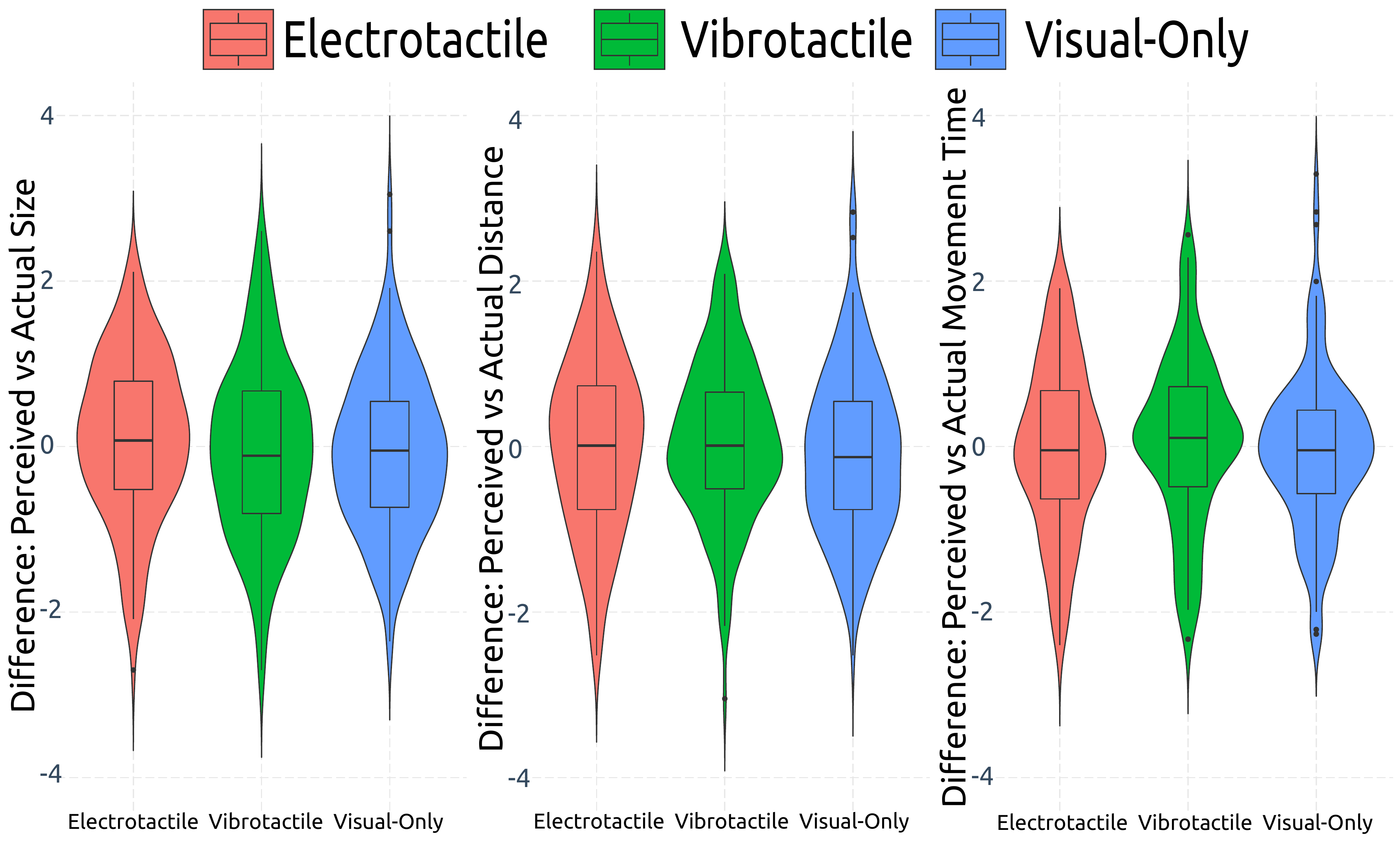}
 \caption{Action-Specific Effects on Perception per Type of Feedback: \\Size (Left), Distance (Centre), and Movement Time (Right).}%
 \label{fig:ASP1}
\end{figure}

\paragraph{Haptic Feedback's Effect:} Discrepantly to \textbf{H3}, the ANOVA analyses for the action-specific effects on perception of target's size $F(1.88,269) = 1.86$, $p =.16$, distance between the targets $F(1.99,285) = 1.61$, $p =.20$, and movement time between the targets $F(1.98,283) = 1.40$, $p =.25$ revealed an absence of a significant main effect of feedback's type on them. Correspondingly, the posthoc comparisons did not reveal any significant difference between the types of feedback. However, as displayed in \autoref{fig:ASP1} and aligned with \textbf{H3}, the patterns of action-specific effects are similar to the observed performance differences amongst types of feedback. In the electrotactile condition, where the participants were more accurate, they also perceived the target larger than its actual size. Also, in the electrotactile and visual-only conditions, where the users were faster, they observed the distance smaller.%

\subsubsection{Regression Analyses}
The accuracy did not correlate with the movement time, and the effect size was diminutive $r=.03, p=.59$, suggesting an absence of interdependence between movement speed and accuracy. However, the index of difficulty was significantly correlated with both accuracy $r=-.35, p<.001$ and movement time $r=.71, p<.001$ with medium and large effect sizes respectively, suggesting less accuracy and greater movement time as the index of difficulty increases. Nevertheless, these associations were expected, since the index of difficulty is defined by the size of targets and the distance between the targets.%

In line with \textbf{H4} \& \textbf{H5}, the action-specific effects on perception were significantly correlated with the performance (see \autoref{tab:Correl}).  Specifically, the improved accuracy was associated with perceiving the target bigger and the movement time longer (i.e., perceiving the time passage slower). Also, faster movement, from the selected target to the next one, was related to perceiving the distance as smaller and the time as shorter (i.e., faster time passage). Lastly, the larger index of difficulty, as defined by the Fitts’s law equation, was correlated with perceiving the target as smaller, the movement time shorter, and the distance between the targets longer. Overall, the effect sizes of the significant correlations with the difference between perceived and actual size, as well as movement time, were of small size. For the associations with the difference between perceived and actual distance, the effect sizes were of medium size.%

Nonetheless, beyond the correlations, which just indicate two-way relationships between the variables, the multivariate regression models postulated a significant impact of the predictors on the action-specific effects (see \autoref{tab:RegModels}). All the models (single predictor + random effects) were significantly better than their respective null models (i.e., including only the random effects), except for the movement time model for the action-specific effect on time perception. Also, the effect sizes of all models were moderate-to-large (e.g., $R^2$ = .22) to extremely large (e.g., $R^2$ = .82). In agreement with \textbf{H4}, the accuracy models appeared as the best (i.e., greater $R^2$ and $\mathcal{\beta}$; variance explained and impact) for target's size perception, as well as time perception. Aligned with \textbf{H5}, the movement time model was seen as the best for the perception of distance between the targets. Finally, given that the responses for movement time were being given after responding for the distance between the targets, we also checked whether the distance difference affected the time difference. The results showed that this model was not better than the null model, and the impact was minuscule ($\mathcal{\beta}$ = 0.01), suggesting that the action-specific effects on time were not affected by the action-specific effects on distance.%

\begin{table}[!h]
\resizebox*{\columnwidth}{1.5in}{%
\fontsize{14pt}{22pt}
\selectfont
\begin{tabular}{@{}cccccc@{}}
\toprule
ASP Variable & Task Variable       & Pearson's r & 95\% CI          & t(430) & p-value            \\ \midrule
\multirow{2}{*}{Difference: Perceived vs Actual Size}          & Accuracy Index & .12  & {[}.03, .22{]}   & 2.07  & .040*              \\
             & Index of Difficulty & -.13        & {[}-.22, -.04{]} & -2.79  & .006**             \\ \hline
\multirow{2}{*}{Difference: Perceived vs Actual Distance}      & Movement Time     & .32  & {[}.23, .40{]}   & 6.92  & \textless{}.001*** \\
             & Index of Difficulty & .36         & {[}.27, .44{]}   & 7.95   & \textless{}.001*** \\ \hline
\multirow{3}{*}{Difference: Perceived vs Actual Movement Time} & Movement Time     & -.19 & {[}-.28, -.10{]} & -4.05 & \textless{}.001***              \\
             & Accuracy Index   & .13        & {[}.04, .22{]} & 2.70  & .007**              \\
             & Index of Difficulty & -.13        & {[}-.23, -.01{]} & -2.82  & .005**              \\ \cmidrule(l){1-6} 
\end{tabular}%
}
\caption{Correlations of Accuracy, Speed, and Index of Difficulty with Action-Specific Perception (ASP) Effects.}
\label{tab:Correl}
\end{table}%

\begin{figure}[!h]
 \centering 
 \includegraphics[width=\columnwidth]{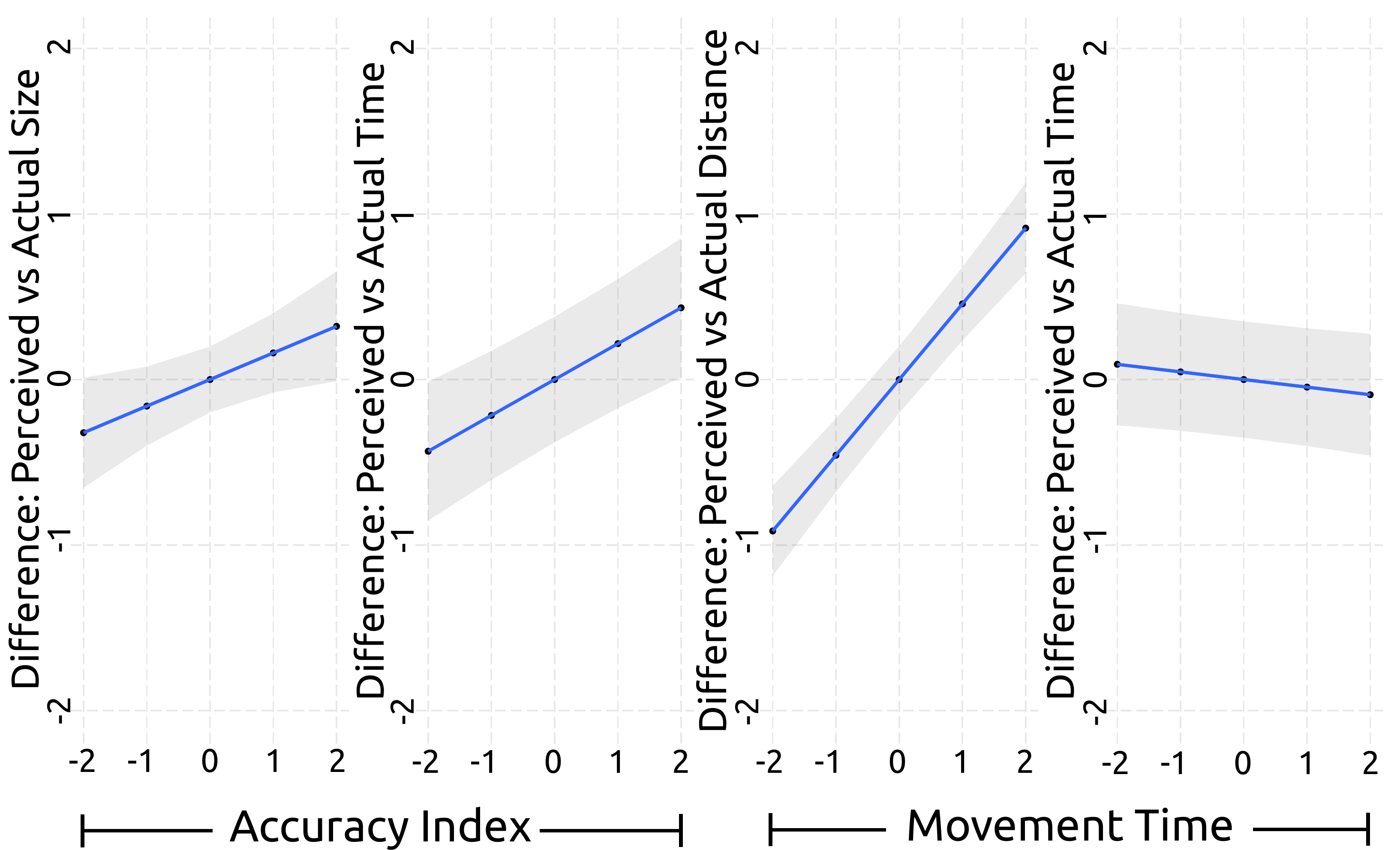}
 \caption{Adjusted Predictions of Action-Specific Effects on Perception: Accuracy on Size Perception (Left), Accuracy on Time Perception (Centre-Left), Movement Time on Distance Perception (Centre-Right), Movement Time on Time Perception (Right).}%
 \label{fig:ASP2}
\end{figure}

\paragraph{Accuracy's Effect on Perception:}For the action-specific effect on the target's size, the accuracy index model was significantly better than the null model, which contained only the random effects per participant as a predictor (see \autoref{tab:RegModels}). The null model explained 19\% of the variance that indicates the implication of individual differences (i.e., random effects per participant) in the action-specific effects on the target's size perception. However, the addition of the accuracy index as a predictor in the respective model explained substantially a larger variance (25\%). Moreover, the accuracy index had the largest $\mathcal{\beta}$ (i.e., the amount of the change of the perception of size, when the accuracy increases by 1 unit), postulating a larger impact on the target's size perception. These results suggest that better accuracy influences the perception of the target as larger than its actual size, while inaccuracy induces to perceive it as smaller (see \autoref{fig:ASP2}). However, the effect of accuracy was even greater on time perception. The null model explained 73\% of the action-specific effect on movement time perception, suggesting the involvement of individual differences. The accuracy model explained a significantly larger amount of the variance (82\%). Similarly, accuracy's $\mathcal{\beta}$ was the largest, postulating an important impact on time perception. The outcomes pinpointed that improved accuracy affects the time passage perception to be slower, while decreased accuracy influences the time passage perception to be faster.%

\begin{table}[!h]
\resizebox*{\columnwidth}{2in}{%
\fontsize{14pt}{22pt}
\selectfont
\begin{tabular}{@{}cccccccc@{}}
\toprule
Predicted &
  Predictor &
  F (1,430) &
  t(430) &
  p (\textgreater{}$|t|$) &
  $\mathcal{\beta}$ coefficient &
  p(\textgreater{}$\mathcal{\chi}^2$) &
  $R^2$ \\ \midrule
\multirow{3}{*}{Difference: Perceived vs Actual Size} &
  Random Effects &
  \textit{NA} &
  \textit{NA} &
  \textit{NA} &
  \textit{NA} &
  \textit{NA} &
  .19 \\
 & Accuracy Index      & 5.62  & 2.37  & .01**              & 0.16  & .01**              & .25 \\
 & Index of Difficulty & 9.56  & -3.09 & \textless{}.001*** & -0.13 & .002**             & .22 \\ \hline
\multirow{3}{*}{Difference: Perceived vs Actual Distance} &
  Random Effects &
  \textit{NA} &
  \textit{NA} &
  \textit{NA} &
  \textit{NA} &
  \textit{NA} &
  .11 \\
 & Movement Time       & 97.3  & 9.86  & \textless{}.001*** & 0.46  & \textless{}.001*** & .40 \\
 & Index of Difficulty & 72.28 & 8.50  & \textless{}.001*** & 0.36  & \textless{}.001*** & .25 \\ \hline
\multirow{4}{*}{Difference: Perceived vs Actual Movement Time} &
  Random Effects &
  \textit{NA} &
  \textit{NA} &
  \textit{NA} &
  \textit{NA} &
  \textit{NA} &
  .73 \\
 & Accuracy Index      & 22.77 & 4.77  & \textless{}.001*** & 0.22  & \textless{}.001*** & .82  \\
 & Movement Time       & 2.72  & -1.65 & .08                & -0.05 & .07                & .75  \\
 & Index of Difficulty & 31.31 & -5.60 & \textless{}.001*** & -0.13 & \textless{}.001*** & .77 \\ \cmidrule(l){1-8}
\end{tabular}%
}
\caption{Multivariate Regression Models of Action-Specific Effects on Perception. The Random Effects model was treated as the Null model. Random effects are included in every model. $\mathcal{\beta}$ = standardized regression coefficient. The p(\textgreater{}$\mathcal{\chi}^2$) pertains to the comparison against the null model. $R^2$x100 = the percentage of explained variance by the model. Based on Cohen \cite{Cohen1992}, $R^2$ = .02 - Small; $R^2$ =.15 - Medium; $R^2$ = .35 - Large.}
\label{tab:RegModels}
\end{table}%

\paragraph{Movement Time's Effect on Perception:} Regarding the prediction of the action-specific effects on the perception of the distance between the targets, the movement time model was substantially improved compared to the null model (see \autoref{tab:RegModels}). The null model explained 11\% of the variance, indicating a moderate effect of individual differences on perceiving the distance. On the contrary, the inclusion of the movement time in the model significantly increased the explained variance (40\%). Also, the movement time has the largest $\mathcal{\beta}$, postulating a substantial impact on the perception of the distance between the targets. Taking them together, these outcomes suggest that the increased speed of movement between two points influences the perception of this distance as smaller. Equally, a slower speed of movement affects the perception of this distance to be longer (see \autoref{fig:ASP2}). Nevertheless, regarding time perception, the movement time model was not significantly better than the null model. While the movement time model explained 75\% of the variance, only the random effects explained 73\%. Finally, the $\mathcal{\beta}$ of movement time was diminutive, indicating a  minor effect on time perception.%

\begin{figure}[!h]
 \centering 
 \includegraphics[width=\columnwidth]{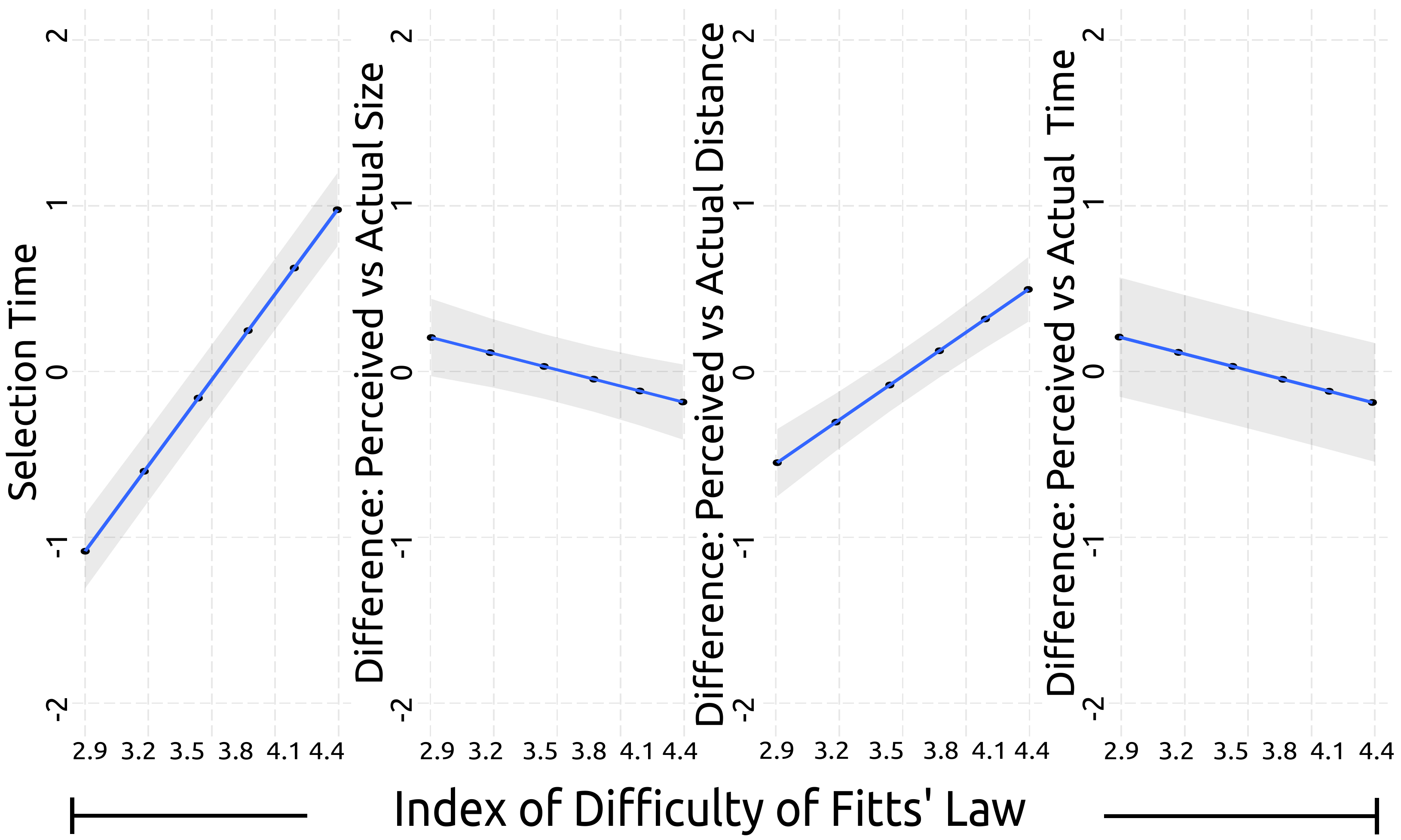}
 \caption{Adjusted Predictions of Index of Difficulty of Fitts’s Law on Performance \& Perception: Selection Time (Left), Size Perception (Centre-Left), Distance Perception (Centre-Right), Time Perception (Right).}%
 \label{fig:Fitts}
\end{figure}

\paragraph{Index of Difficulty on Perception:} The index of difficulty of Fitts’s law efficiently explained 86\% of the variance of performance (i.e., selection time) on the task, validating the expected increase of selection time (i.e., slower completion) as the difficulty incrementally augments (see \autoref{fig:Fitts}). Interestingly, the index of difficulty was also a significant predictor of the action-specific effects on perception. The index of difficulty’s models were all significantly better than the corresponding null models. The respective index of difficulty model explained a substantial amount of the variance of action-specific effects on the perception of the target's size (22\%), the distance between the targets (25\%), and movement time (77\%; see also \autoref{fig:Fitts}), albeit that none of these models was the one which explained the greatest amount of variance. Comparably, the $\mathcal{\beta}$ of the index of difficulty indicated an important impact on the perception of the target's size, the distance between the targets, and movement time (see \autoref{tab:RegModels}). In agreement with \textbf{H5}, these findings postulate that the increase of difficulty influences perceiving the target smaller, the distance longer, and the time shorter (see \autoref{fig:Fitts}).%

\section{Discussion}
\subsection{Performance}
Selection and Reaction time was found to be substantially slower in visuo-vibrotactile feedback condition. These findings are in agreement with the findings in the study of Bricker \emph{et al.,} \cite{Brickler}, where the visuo-haptic condition showed significantly slower selection times on a Fitts’s law task in VR. Also, our findings align with the observation of Ariza \emph{et al.,} \cite{Ariza}, where the vibrotactile showed longer correction phases on Fitts’s law task in VR. Considering that there are several discrepancies amongst studies which had used non-standardized tasks, we may infer that the examination of the effects of tactile modalities on performance can be assessed more reliably by implementing standardized tasks. This methodological approach may hence alleviate the frequency of producing discrepant outcomes.%

Furthermore, the accuracy in visuo-vibrotactile condition was found to be considerably lower than in visual condition. This finding agrees with the observations in previous studies \cite{Cheng,Cholewiak}. However, these studies implemented different non-standardized tasks. We may infer that the agreement among the observations in three different tasks postulates that the reason for decreased accuracy by vibrotactile feedback is not task-specific. A possible reason could be the uncanny haptic effect that occurs in incongruent bi- or multi-modal feedbacks \cite{Berger}. However, this adverse effect on performance should be further explored on both standardized and non-standardized tasks, and by using diverse vibrotactile devices.%

Comparably to the previous studies (i.e., \cite{Pamungkas2, Sagardia, Hummel,Vizcay2021}), electrotactile feedback was found to facilitate improved accuracy in VR. However, our findings were observed on a standardized target acquisition (Fitss’s law) task. Notably, the observed improvement in accuracy was detected against the visuo-vibrotactile and visual feedback.  In our study, the accuracy in visuo-electrotactile condition was significantly better than in \ visual-only and visuo-vibrotactile conditions. However, the effect sizes were not moderate-to-large like in the study of Vizcay \emph{et al.,} \cite{Vizcay2021}. This may be attributed to the fact that the electrotactile feedback was binary, and not dynamic as in their study. The implementation of dynamic electrotactile feedback could have enabled a further improved accuracy, which could have a large effect size against other feedback modalities. Nonetheless, this should be examined in future endeavours.%
Finally, the self-reports of participants on the utility of electrotactile feedback, in terms of accuracy, alarming, and salience, agreed with the observed gains in accuracy. However, the self-reports showed a comparable utility of vibrotactile in the same terms, while the accuracy in visuo-vibrotactile condition was substantially worse than in visual and visuo-electrotactile conditions. This disagreement between objective (measured performance) and subjective (self-report) outcomes may be attributed to the fact that vibrotactile feedback is a very familiar sensation which has broadly been used by the participants. This finding also shows the significant effect of subjectivity on self-reports. Comparably to Slater's recommendations for investigating presence in VR \cite{Slater}, the self-reports on performance should be only considered as complementary to the objective metrics.%

\subsection{ASP}
The El Greco evaluations of size, distance, and time were all significantly different from the initial evaluations. Also, the direction of the differences in El Greco evaluation was always towards the actual size, distance, or time. This outcome confirms that the observed effects were indeed action-specific effects on spatial and time perception, and not a by-product of response bias. Nevertheless, the effect size of the difference between the initial and El Greco evaluation of time perception was not as large as the effect size in size and distance perception. This may be a result that for time perception was not any explicit feedback besides the change of target's size and distance between them. In future endeavours, explicit temporal feedback should be offered to the participants during the El Greco evaluation to achieve effect sizes comparable to the size and distance perception.%

Although performance increases and decreases per feedback condition were reflected in the action-specific effects in each feedback condition, the differences between the mediated action-specific effects were not significant. This absence of significant differences, regarding action-specific effects on perception, may be attributed to the small effect sizes of performance differences, albeit significant. This outcome agrees with the argumentation of Dragicevic \cite{Dragicevic2016} that effect sizes are of augmented importance in HCI, compared to p-values. In this study, due to the inability of the vibrotactile actuator to deliver dynamic feedback without a perceivable latency, a binary approach was adopted for every feedback condition. Previously, dynamic electrotactile \cite{Vizcay2021} and visual \cite{Ariza} feedback had demonstrated moderate and large effect sizes. Thus, the implementation of dynamic feedback types may detect significant differences in action-specific effects. Though, this has to be examined in future experiments.%

In support of the ASP theory, we found significant effects of performance and task difficulty on size, distance, and time perception. In line with \cite{Witt2005Ball} increases in accuracy influence participants to perceive the target's size as larger, while decreases in accuracy influence participants to perceive it as smaller. Furthermore, aligned with \cite{Witt2010Tennis,Witt2017}, increased accuracy modulated a slower time passage, while decreased accuracy in a faster one. Also, faster hand movements affected participants to perceive the distance as shorter, while slower hand movements affected participants to perceive it as longer. This finding is in accordance with the findings in chronic pain patients \cite{witt2009long}, hikers \cite{Proffitt2003}, and individuals with extra weight \cite{SUGOVIC2016}, as well as the uphill experiment in VR \cite{Laitin2019}. However, these studies were not focused on the effects of performance (i.e., movement speed) on perception, but rather on the task's difficulty effect on perception.%

Indeed, our findings postulated that the objective metric of Fitts’s  law's index of difficulty had an important impact on the size, distance, and time perception. Our findings are in utter accordance with the outcomes of previous studies on the effect of task difficulty on distance perception \cite{Laitin2019,Proffitt2003,SUGOVIC2016}, as well as size and time perception \cite{Witt2017}. Our study has hence highlighted a new dimension of the index of difficulty of Fitts’s law. Beyond performance, the index of difficulty can therefore be used for assessing and predicting perception, since it asserts an effect on task-related perceptual processes.%

As mentioned earlier, the ASP is significant when the intention to act is present \cite{Witt2010}. In VR and HCI, we would call this state, the interactability of the components of a virtual environment with the user. Since interaction is the cornerstone of VR applications and studies, the action-specific effects of performance or task difficulty on perception should be considered. ASP can contribute in several aspects, such as the design of user interfaces and the personalization of VR experiences.  For example, based on ASP, when a user is highly accurate in pressing the buttons of a cluttered user interface, the user perceives the targets/buttons as larger. Thus, the targets/buttons can be further decreased in size (e.g., proportionally to the effective width) which can either decrease the clutteredness of the user interface or increase the quantity of the available targets of the user interface. Since ASP changes within an individual (e.g., when the user is faster, then distances are perceived shorer; when the user is slower, then the distances are perceived longer), a machine learning algorithm can be used to detect these changes in performance and modulate the components of the virtual environment respectively (e.g., increase decrease distances and sizes of the targets).%

As mentioned above, the effect sizes are really important in HCI~\cite{Dragicevic2016}. Since the magnitude of ASP was found to be modulated by the effect sizes, the ASP can be used as a parsimonious evaluation of performance improvements, and input devices' efficiency. In our study, while the binary electrotactile feedback was found to facilitate significantly better accuracy, yet with a small size effect, this difference was not replicated in the ASP, which led us to infer that dynamic feedback could have been a better option to increase the magnitude of the effect. Similarly, beyond just detecting significant differences, the ASP can be used to parsimoniously assess the impact of other input devices (e.g., VR gloves) or techniques (e.g., gaze interaction) on improving users' performance. %

Our findings bring also implications to the study of presence and embodiment. One of the illusions that enhances the sense of presence in VR is the plausibility illusion \cite{Slater2}. The intention to act and the performance of inter-actions are central to both plausibility illusion and the ASP. Thus, the ASP (e.g., stronger action-specific effects on perception) could be an indicator of how strong the plausibility illusion is, which also indicates that the user experiences a stronger sense of presence. Future VR studies should attempt to scrutinize this overlap and the potential relationship between ASP, plausibility illusion, and presence.%

Similarly, in embodiment studies, researchers have seen perceptual alterations that have been attributed to the strength of the embodiment (e.g., \cite{van2014,Linkenauger2015}). For example, Banakou \emph{et al.,} \cite{Banakou} found that the embodiment of a child-like avatar modulates an overestimation of the objects. However, the participants were interacting with these objects. Considering our results and ASP theory, the observed over-estimation could have been attributed to or modulated by an action-specific effect on size perception due to the increased difficulty to interact with the objects, while having child-size hands.  Thus, we cannot conclude whether this is purely attributed to the strength of embodiment illusion, ASP, or the interaction between the two. The effects of ASP during embodiment, as well as the interaction between embodiment and ASP, should also be examined in future VR studies on the embodiment illusion.%

\subsection{Limitations}
The present study also suffers from certain limitations. The implemented visuo-electrotactile feedback was binary and not dynamic, while previous studies indicated the performance gains of dynamic electrotactile feedback. As a result, while we observed statistically significant improvements in accuracy, the magnitude of these improvements was small.  Future studies should attempt to investigate the performance benefits of uni- or bi-modal dynamic electrotactile feedback on standardized tasks in VR. Furthermore, there was no explicit feedback on movement time during the El Greco evaluation of action-specific effects on time perception. Despite that the difference between the initial and the El Greco evaluation of time perception was statistically significant, the effect size of this difference was small. Given that the effect sizes of the differences between the evaluations of size and distance were large, future studies should attempt to provide explicit feedback on time, to efficiently achieve effect sizes comparable to the spatial perception.%

\section{Conclusion}
The electrotactile was identified as the feedback type that facilitates the best accuracy while having reaction times comparable to visual feedback. However, the improvement in accuracy was of a small size. While the performance differences of feedback types were reflected in ASP, the differences were not significant, which postulates that the ASP is a more parsimonious method to evaluate the impact of haptic and input devices on performance. However, performance (accuracy and speed) and task difficulty had significant effects on size, distance, and time perception. Our research outcomes suggest that the user's ability, in interacting with the components of a virtual environment, affects how the virtual environment is perceived. Our findings also surfaced a new utility of the index of difficulty of Fitts’s law for predicting and explaining action-specific effects on perception. ASP should be considered in future VR studies, especially for systems evaluation, sense of presence, and embodiment. Future studies should strive to scrutinize further the role of ASP in VR and its relationships with user performance in diverse scenarios, and aspects of presence such as plausibility and embodiment illusions.%


\acknowledgments{
This work was supported by the European Union’s Horizon 2020 research and innovation program under grant agreement No. 856718 (TACTILITY).
}

\bibliographystyle{abbrv-doi}

\bibliography{my}
\end{document}